\documentclass{raa}

\usepackage{graphicx,epsfig,epsf,amssymb}
\usepackage{graphicx,times}             

\begin{document}

   \title{$^{12}$CO $J$=2-1 and $^{12}$CO $J$=3-2  Observations toward the High-Mass Protostellar Candidate IRAS 20188+3928
}

   \volnopage{{\bf 2013} Vol. {\bf 13} No.0, 000--000}      
   \setcounter{page}{1}          

   \author{Jin-Long.  Xu
      \inst{1,2}
   \and Jun-Jie Wang
      \inst{1,2}
   }

   \institute{National Astronomical Observatories, Chinese Academy of Sciences,
             Beijing 100012, China {\it xujl@bao.ac.cn}\\
        \and
             NAOC-TU Joint Center for Astrophysics, Lhasa 850000, China\\
\vs\no
   {\small Received~~2012 July 2; accepted~~2012~~July 26}}

\abstract{We have carried out $^{12}$CO $J$=2-1 and $^{12}$CO
$J$=3-2 observations toward the high-mass protostellar candidate
IRAS 20188+3928. Comparing with the previous observations, the
$^{12}$CO $J$=2-1 and $^{12}$CO $J$=3-2 lines both have asymmetric
profiles with an absorption dip. The velocity of the absorption dip
is $\sim$ 1.0 km $\rm s^{-1}$. The spectral shape may be caused by
rotation. The velocity-integrated intensity map and
position-velocity diagram of the $^{12}$CO $J$=2-1 line present the
obvious bipolar component, further verifying this region has an
outflow motion. This region also is associated with an HII region,
an IRAS source, and H$_{2}$O maser.  The H$_{2}$O maser has the
velocity of 1.1 km s$^{-1}$ . Comparing with the components of the
outflow, we find that the H$_{2}$O maser is not associated with the
outflow. Using the LVG model, we obtained that possible averaged gas
density of the blueshifted lobe and redshifted lobe are
1.0$\times$10$^{5}$ cm$^{-3}$ and 2.0$\times$10$^{4}$ cm$^{-3}$,
while kinetic temperature are 26.9 K and 52.9 K, respectively.
Additionally, the outflow has the higher integrated intensity ratio
($I_{\rm CO}$$_ {J=3-2}$/$I_{\rm CO}$$_ {J=2-1}$).   }

   \authorrunning{J.-L. Xu, \& J. J. Wang}            
   \titlerunning{$^{12}$CO $J$=2-1 and $^{12}$CO $J$=3-2 observations toward IRAS 20188+3928}  

   \maketitle

%
%
\section{Introduction}           
\label{sect:intro} IRAS 20188+3928 is a high-mass protostellar
candidate (Zhang et al . \cite{zhang05}), located in the Cygnus
region at the position ($l, b$) = (77.46, 1.76) (Little et al.
\cite{little88}). The distance estimated to this source varies from
0.31 to 4 kpc (Little et al. \cite{little88}; Palla et al.
\cite{palla91}; Molinari et al. \cite{molinari96}), hence, the
luminosity is highly uncertain. A bipolar molecular outflow was
discovered in HCO$^{+}$ $J$=2-1 and $^{13}$CO $J$=1-0 emission lines
in the IRAS 20188+3928 region (Little et al. 1988). They suggested
that the outflowing gas has a dense and clumpy nature in the
north-SW direction, and IRAS 20188+3928 is not obviously associated
with a well known stellar cluster and HII region. Zhang et al.
(\cite{zhang05}) observed the bipolar outflow in $^{12}$CO $J$=2-1
centered on the IRAS source and roughly in the NS direction, but
their observation can not fully cover the emission of the bipolar
outflow. In addition, Varricatt et al. (\cite{varricatt10}) not only
detected the bipolar outflow in H$_{2}$ line, but also an additional
outflow in the region. Because the outflows in this region are
complicated, farther observations are needed toward the IRAS
20188+3928 region.

Additionally, IRAS 20188+3928 is associated with a compact molecular
cloud (Anglada et al. \cite{anglada97}). An H$_{2}$O maser emission
was detected by several other investigators (Palla et al.
\cite{palla91}; Brand et al. \cite{brand94}; Jenness et al.
\cite{jenness95}). Several observers detected NH$_{3}$ emission
(Molinari et al. \cite{molinari96}; Anglada et al. \cite{anglada97};
Jijina et al. \cite{jijina99}), implying that the IRAS source is
deeply embedded in the high density gas, which may be the driving
source of the bipolar outflow. Jenness et al. (\cite{jenness95})
imaged two sources at 450 and 800 ¦Ìm, agreeing with the position of
two 6-cm radio emission sources detected by Molinari et al.
(\cite{molinari98}), the brighter and the fainter ones located
north-west of the IRAS position by 4.2 and 28.4 arcsec,
respectively. Near-infrared polarimetry revealed an illuminating
source, about 6$^{\prime\prime}$ north of the IRAS source (Yao et
al. \cite{yao00}). This illuminating source is better centered on
the CO outflow than the IRAS source. Accordingly, near-IR images
show a cluster of deeply embedded objects towards the centre of the
IRAS 20188+3928 region, very close to the IRAS 20188+3928 position.
The agreement of these positions indicates that the outflow source
is located in the cluster. Hence, there are multiple YSOs in the
region, the driving source of the bipolar outflow have not been well
identified.

In this paper, we have carried out the $^{12}$CO $J$=2-1 and
$^{12}$CO $J$=3-2 observations towards the high-mass protostellar
candidate IRAS 20188+3928. Clearly north-SW outflow and possible
rotation may be associated with IRAS 20188+3928.

\section{Observations}
\label{sect:Obs} The mapping observations of IRAS 20188+3928 were
performed in $^{12}$CO $J$=2-1 and $^{12}$CO $J$=3-2 lines using the
KOSMA 3m telescope at Gornergrat, Switzerland,  in November 2009.
The half--power beam widths of the telescope at the observing
frequencies of 230.538 GHz and 345.789 GHz are $130^{\prime\prime}$
and  $80^{\prime\prime}$, respectively. The pointing and tracking
accuracy was better than 20$^{\prime\prime}$. The medium and
variable resolution acousto optical spectrometers have 1501 and 1601
channels, with total bandwidths of 248 MHz and 544 MHz. The channel
widths of 165 and 340 kHz correspond to velocity resolutions of 0.21
and 0.29 ${\rm km\ s^{-1}}$, respectively. The beam efficiency
$B_{\rm eff}$ is 0.68 at both 230 GHz and 0.72 at 345 GHz. The
forward efficiency $F_{\rm eff}$ is 0.93 for all frequencies.
Mapping observations were centered at RA(J2000)=$20^{\rm h}20^{\rm
m}39.30^{\rm s}$, DEC(J2000)=$39^{\circ}37'51.90^{\prime\prime}$
using the on-the-fly mode, the total mapping area is
$11^{\prime}\times 11^{\prime}$ with a $1^{\prime}\times 1^{\prime}$
grid. The correction for the line intensities to {the }main beam
temperature scale was made using the formula $T_{\rm mb}= (F_{\rm
eff}/B_{\rm eff}\times T^\ast_{\rm A})$. The data were reduced using
the GILDAS/CLASS \footnote{http://www.iram.fr/IRAMFR/GILDAS/}
package.

The 1.4GHz radio continuum emission data were obtained from the NRAO
VLA Sky Survey (NVSS; Condon et al. 1998).

\section{Results and Discussion}
\label{sect:data}
\subsection{CO molecular spectra}
Fig. 1 shows the spectra of $^{12}$CO $J$=2-1, $^{12}$CO $J$=3-2,
and $^{13}$CO $J$=1-0 at the IRAS 20188+3928 position. The $^{13}$CO
$J$=1-0 data is obtained from the Purple Mountain Observatory (PMO)
archive data\footnote{ http://www.radioast.csdb.cn}. Each spectrum
display broad line wings. $^{12}$CO $J$=2-1 and $^{12}$CO $J$=3-2
lines present asymmetric profiles with double peaks, while $^{13}$CO
$J$=1-0 line shows a single peak profile.  Wu et al. (\cite{wu10})
only detected one compact core from the IRAS 20188+3928 region in
the HCN $J$=1-0 and CS $J$=2-1 lines, suggesting that there is one
velocity component. The profiles of $^{12}$CO $J$=2-1 and $^{12}$CO
$J$=3-2 with double peaks may be a indication of absorption. The
velocity of an absorption dip is $\sim$ 1.0 km $\rm s^{-1}$.
Rotation can be responsible for the spectral shape. With the
single-dish data currently available, we did not find any kinematic
evidence of rotation.  The $^{13}$CO $J$=1-0 line is optically thin,
which can be used to determine the systemic velocity. A systemic
velocity of $\sim$ 2.0 km $\rm s^{-1}$ is obtained from this line.
According to the Galactic rotation model of Fich et al.
(\cite{fich89}) together with $R_{\odot}$ = 8.5 kpc and $V_{\odot}$
= 220 km s$^{-1}$, where $V_{\odot}$ is the circular rotation speed
of the Galaxy, we obtain a kinematic distance of ~0.32 kpc to IRAS
20188+3928, which is used to calculate the physical parameters in
this paper. The full widths (FW) of the $^{12}$CO $J$=2-1 and
$^{12}$CO $J$=3-2 lines both are about 20 km $\rm s^{-1}$ from -10
km $\rm s^{-1}$ to 10 km $\rm s^{-1}$. The large FW appear a strong
indication of outflow motion.

   \begin{figure}[h!]
   \centering
   \includegraphics[width=8cm, angle=180]{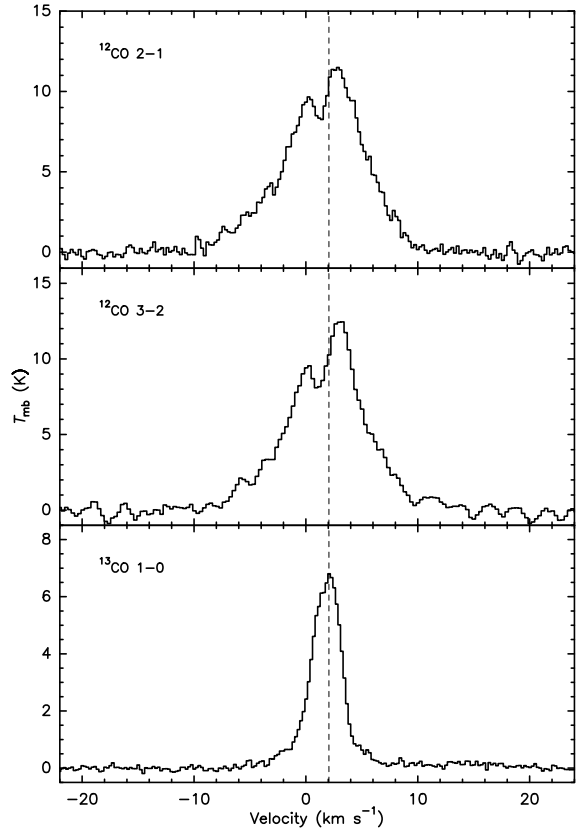}
   \vspace{-8mm}
   \caption{Spectral profiles observed at the central
position of IRAS 20188+3928 in the optically thick $\rm ^{12}$CO
$J$=2-1, $\rm ^{12}$CO $J$=3-2 lines, and the optically thin $\rm
^{13}$CO $J$=1-0 line. The dotted line in the spectra marks the
cloud systemic velocity.}
   \label{Fig:demo1}
\vspace{-8mm}
   \end{figure}

\subsection{ The bipolar outflow}
\subsubsection{ The outflow morphology}
To determine the velocity components and morphology of the outflow,
we made a position-velocity (PV) diagram with a cut along the
north-south direction as shown in Fig. 2. The PV diagram in Fig. 2
clearly shows bipolar components, but we cannot see the emission
from the core of IRAS 20188+3928, indicating that above cut
direction represents that of the bipolar components. The blueshifted
and redshifted components have obvious velocity gradients from
$-$7.6 to $-$0.6 km $\rm s^{-1}$ and 3.2 to 9.0 km $\rm s^{-1}$,
respectively. The distributions of redshifted and blueshifted
velocity components in Fig.2 further confirm a bipolar outflow in
this region. Using the velocity ranges of the blueshifted and
redshifted components, we made the integrated intensity map as shown
in Fig. 3. In Fig. 3, the blue and red contours represent the
blueshifted and redshifted components of the outflow. The outflow
clearly displays the north-SW components, which is associated with
observation in HCO$^{+}$ $J$=2-1 and $^{13}$CO $J$=1-0 emission
lines (Little et al. 1988). The 1.4 GHz continuum emission is
superimposed on the outflow, which may come from an HII region
(Urquhart et al. \cite{Urquhart09}). Both HII region and infrared
source IRAS 20188+3928 are located the beginning position of the
outflow, because of our lower spatial resolution we cannot identify
which object is the driving source of the outflow. An H$_{2}$O maser
was detected in this region, which has the velocity of 1.1 km
s$^{-1}$ (Palla et al. \cite{palla91}; Brand et al. \cite{brand94};
Jenness et al. \cite{jenness95}). Comparing the components of the
outflow, we find that the H$_{2}$O maser is not associated with the
outflow, but with the core. If an accretion motion is further
confirmed in this region, then the H$_{2}$O maser may come from the
disk.
   \begin{figure}
   \centering
   \includegraphics[width=8cm, angle=0]{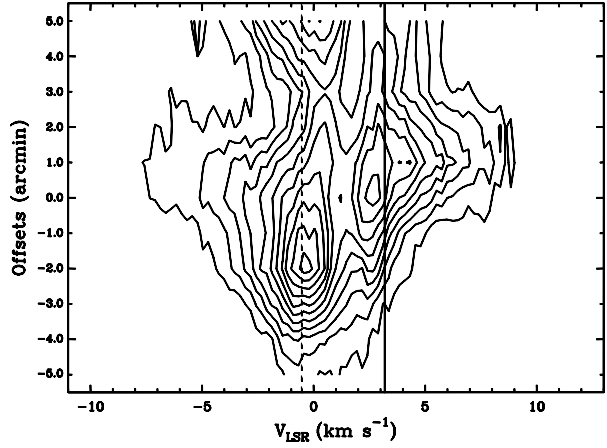}
   \caption {P-V diagram constructed from the $^{12}$CO $J$=2-1 transition. Contour levels are 1, 2, 3, 4.5, 6, 6.5, 7.5,
  9, 10.5, 12, 14,..., 25 K. The vertical dashed and solid lines indicate
  the beginning of the blueshifted and redshifted emission, respectively. }
   \label{Fig:demo1}
   \end{figure}

   \begin{figure}[h]
   \centering
   \includegraphics[width=9cm, angle=270]{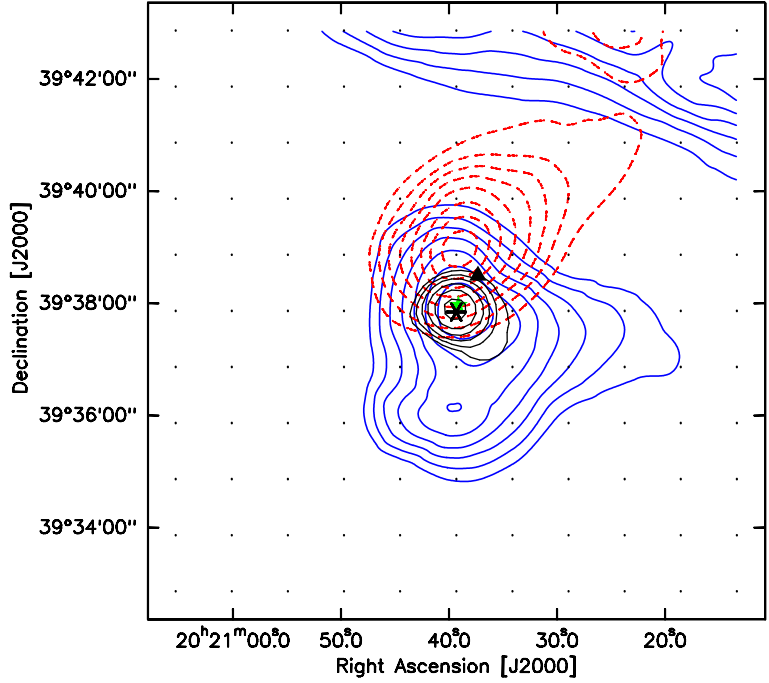}
   \caption{The velocity-integrated intensity map of
$^{12}$CO $J$=2-1 outflow (red and blue contours) overlaid with the
1.4 GHz emission (black contours). The black contour levels are at
3, 6, 9, 12, 15, 18, and 21 $\sigma$ (1 $\sigma$ is 0.002 $\rm Jy\
beam^{-1}$).  The red and blue contour levels are 35, 50,..., 95
$\%$ of the peak value.  The $\rm H_{2}O$ maser is shown by a filled
triangle. The ``$\ast$'' symbols indicates the position of IRAS
20188+3928 and HII region.}
   \label{Fig:demo1}
   \end{figure}

\subsubsection{ The outflow parameters}
In this section, we use a large velocity gradient (LVG) radiative
transfer model (Goldreich et al. \cite{Goldreich}; Qin et al. \cite{qin08a}) to determine the gas density ($n(\rm H_{2})$) and
kinetic temperature ($T_{\rm kin}$) of the outflow. The blueshifted
and redshifted velocity intervals ($\Delta v$) are from $-$7.6 to
$-$0.6 km $\rm s^{-1}$ and from 3.2 to 9.0 km $\rm s^{-1}$,
respectively. In this model, assuming a uniform density and using
above velocity intervals, $n(\rm H_{2})$ and $T_{\rm kin}$ are
obtained by fitting the line intensity of $^{12}$CO $J$=2-1 and the
line ratio of $^{12}$CO $J$=3-2/$^{12}$CO $J$=2-1, which are varied
in a reasonable range for this region. Fig 4. shows the results of
the LVG modeling. From the Fig.4,  various column densities are from
3.5$\times$10$^{4}$ to 1.7$\times$10$^{5}$ cm$^{-3}$ with
corresponding $T_{\rm kin}$ ranging from 22.9 to 30.9 K for the
blueshifted lobe. The average values of possible $n(\rm H_{2})$ and
$T_{\rm kin}$ are 1.0$\times$10$^{5}$ cm$^{-3}$ and 26.9 K,
respectively. For redshifted lobe, $n(\rm H_{2})$ ranges from
1.6$\times$10$^{4}$ to 2.8$\times$10$^{4}$ cm$^{-3}$, while $T_{\rm
kin}$ is from 39.8 to 83.1 K. The averaged $n(\rm H_{2})$ and
$T_{\rm kin}$ are 2.0$\times$10$^{4}$ cm$^{-3}$ and 52.9 K,
respectively.

Additionally, if both lobes are approximately spherical in shape,
their mass is given by $M$=$n(\rm H_{2})$$\frac{1}{6}\pi
L^{3}\mu_{\rm g}m(\rm H_{2})$ (Garden et al. \cite{Garden}), where
$\mu_{\rm g}$=1.36 is the mean atomic weight of the gas, $L$ is the
lobe diameter , and $m(\rm H_{2})$ is the mass of a hydrogen
molecule. We calculate the dynamic timescales using $t_{\rm d}=r/v$,
where $v$ is the maximum flow velocity relative to the cloud
systemic velocity , and $r$ is the length of the begin-to-end flow
extension for each lobe. The mass entrainment rate of the outflow is
determined by $\dot{M}=M/(t_{\rm d})$. The momentum $P$ and energy
$E$ are calculated by $P=MV$ and $E=MV^{2}$, where $V$ is the mean
velocity of the gas relative to the cloud systemic velocity. The
physical parameters and the calculated results of the outflow are
listed in Table 1. A total mass and mass entrainment rate of the
outflow are 3.5 $\rm M_{\odot}$ and 6.2 $\times$ $10^{-5}$ $\rm
M_{\odot}\ yr^{-1}$, respectively. The average dynamical timescale
is about 6.2 $\times$ $10^{4}$ yr.
   \begin{figure}
   \vspace{0mm}
   \centering
   \includegraphics[width=7.5cm, angle=270]{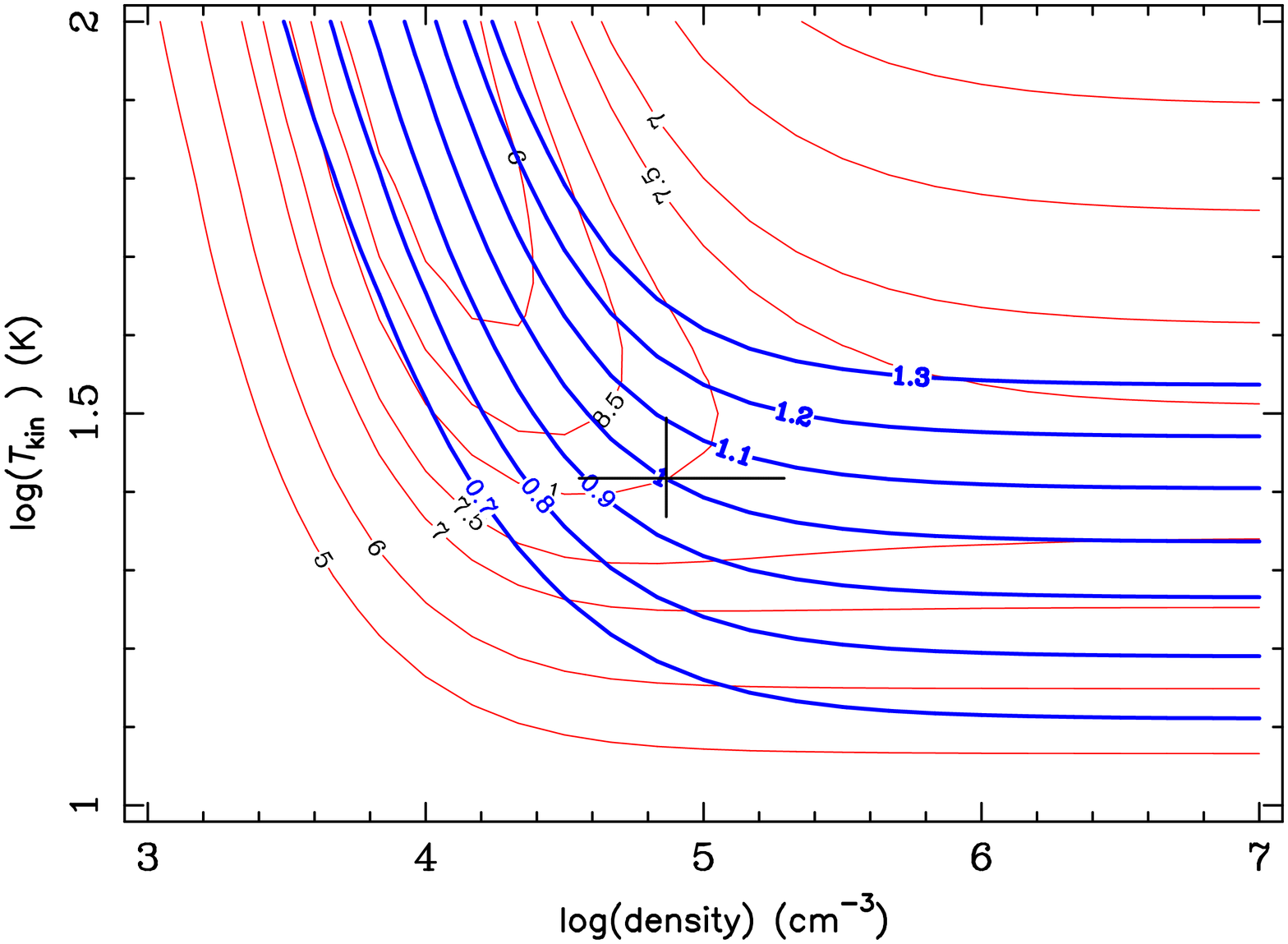}
   \end{figure}
   \begin{figure}
   \vspace{0mm}
   \centering
   \includegraphics[width=7.5cm, angle=270]{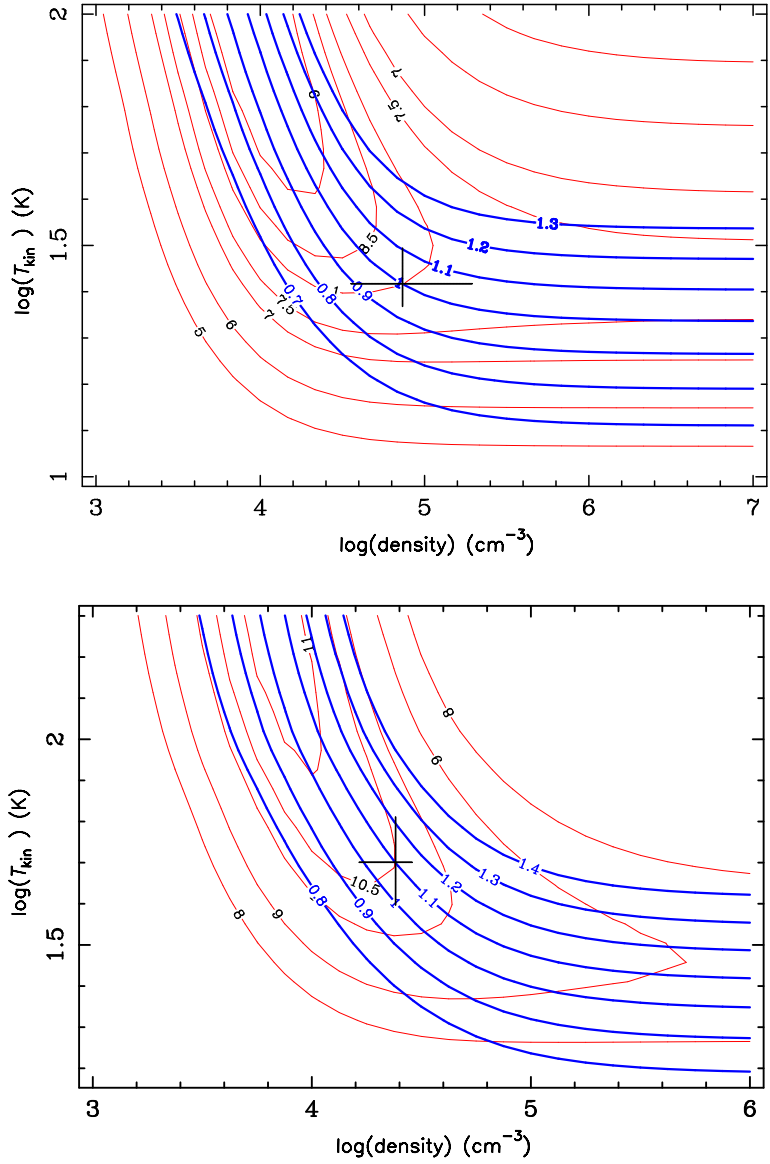}
   \vspace{0mm}
   \caption{LVG model analysis using the $^{12}$CO $J$=2-1 and $^{12}$CO $J$=3-2
lines. $Top$: The blueshifted components of the outflow. $Bottom$:
The redshifted components of the outflow. The horizontal and
vertical axes are the volume density $n_{\rm H_{2}}$ and kinetic
temperature $T_{\rm kin}$ in log scale, respectively. Blue curves
represent line intensity of $^{12}$CO $J$=2-1, while red curves
represent the line ratios of CO($J$=3-2)/CO($J$=2-1). The results of
$T_{\rm kin}$ and $n_{\rm H_{2}}$ are taken from the cross points of
red and blue curves. The error bars indicate the range of possible
solutions. }
   \label{Fig:demo1}
   \end{figure}

\begin{table}[]
\tabcolsep 2.5mm\caption[]{ The physical parameters of the outflow.
}\vspace*{-25pt} \vspace{3mm}
  \label{Tab:publ-works}
  \def\temptablewidth{-6\textwidth}
  \begin{center}\begin{tabular}{cccccccccccccccc}
  \hline\noalign{\smallskip}
  \hline\noalign{\smallskip}
Wing & $T_{\rm kin}$ & $n(\rm H_{2})$& $M$  & $t_{d}$ &$\dot{M}$&$P$&$E$           \\
        &\rm K & ($\rm \times10^{5}cm^{-3}$)&  ($\rm M_{\odot}$)& ($\rm \times10^{4}yr$) &($\rm M_{\odot}\ yr^{-1}$) &($\rm M_{\odot}\ km\ s^{-1}$)  &($\rm \times10^{44}erg$)           \\
  \hline\noalign{\smallskip}
Blue  & 26.9 & 1.0 & 3.3 & 5.6& $5.9\times10^{-5}$  & 11.6 & 4.0  \\  
Red   & 52.9 & 0.2 & 0.2 & 6.8& $0.3\times10^{-5}$  & 0.6  & 0.2 \\
\noalign{\smallskip}\hline
\end{tabular}\end{center}
\vspace{0mm}
\end{table}

\subsubsection{ The integrated intensity ratio ($I_{\rm CO}$$_ {J=3-2}$/$I_{\rm CO}$$_
{J=2-1}$)}

The line intensity ratios based on the optically thick CO
transitions can indicate the temperature varies at different
positions (Qin et al. \cite{qin08b}), and further trace shocks  (Xu
et al. \cite{xua,xub,xu12}). When the outflow comes into contact
with the surrounding interstellar medium (ISM), this process can
product the shock. In order to obtain the integrated intensity ratio
of $^{12}$CO $J$=3-2 and $J$=2-1 ($I_{\rm CO}$$_ {J=3-2}$/$I_{\rm
CO}$$_ {J=2-1}$), we convolved the 80$^{\prime\prime}$ of $^{12}$CO
$J$=3-2 data with an effective beam size of
$\sqrt{130^{2}-80^{2}}=102^{\prime\prime}$. The integrated
intensities were calculated for $^{12}$CO $J$=2-1 line in the same
velocity range as for $^{12}$CO $J$=3-2. Fig.5 shows the
distribution of the ratio (color scale) overlaid with the
distribution of the $^{12}$CO $J$=2-1 line integrated intensity of
the outflow (black solid and dashed contours). From the Fig.5, we
can see that the distribution of the ratios both presents shell-like
morphology. The maximum ratio value is 1.2 in the blueshifted lobe,
while the redshifted lobe has the maximum ratio value of 1.7. The
rms level is  0.10 (1$\sigma$). The molecular clouds associated with
HII regions have higher ratio and have higher gas temperature than
those sources without HII regions, indicating that the high line
ratio may be due to heating of the gas by the massive stars (Wilson
et al. 1997). While the high ratio values in the molecular clouds
interacting with SNR (Xu et al. \cite{xua,xub,xu12}) exceed previous
measurement of individual Galactic molecular clouds, implying that
the SNR shock has driven into the molecular clouds. Here, the high
ratio values detected in the region of the outflow may be caused by
the outflow interacting with the surrounding ISM. In addition,  the
maximum ratio value of the redshifted lobe is greater than that of
the blueshifted lobe, we suggest that the redshifted lobe has the
higher temperature, which is associated the results derived from the
LVG model. In section 3.2.2, we obtain that the redshifted lobe has
the kinetic temperature of 52.9 K, which is larger than that (26.9
K) of the blueshifted lobe.

   \begin{figure}
   \vspace{-28mm}
   \centering
   \includegraphics[width=13cm, angle=270]{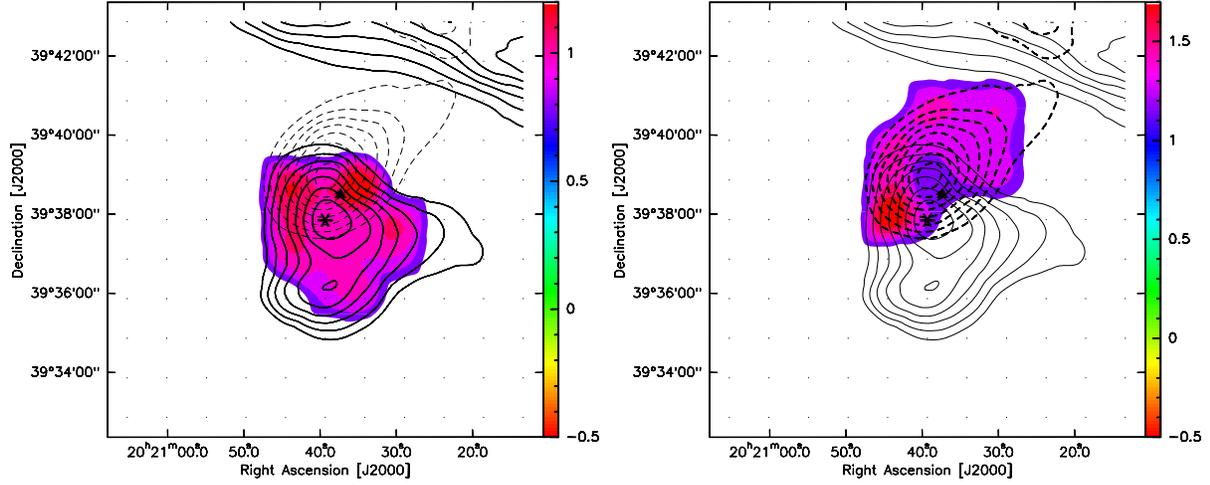}
   \vspace{-38mm}
   \caption{$^{12}$CO $J$=2-1 intensity maps of the outflow are superimposed on the line
intensity ratio maps (color scale). The wedge indicates the
line-intensity ratio scale. Left panel: the blueshifted lobe; Right
panel: the redshifted lobe.}
   \label{Fig:demo1}
   \end{figure}
\section{Conclusions}
\label{sect:conclusion} We present  submillimeter/millimeter
observations of the high-mass protostellar candidate IRAS 20188+3928
in the $\rm ^{12}$CO $J$=2-1 and $\rm ^{12}$CO $J$=3-2 lines. The
spectra profiles of the $^{12}$CO $J$=2-1 and $^{12}$CO $J$=3-2
both with an absorption dip, together with the $^{13}$CO $J$=1-0
line with a single peak profile, indicate that possible rotation may
be associated with IRAS 20188+3928. The $\rm ^{12}$CO $J$=2-1
velocity-integrated intensity map, position-velocity diagram, and
the broad wing (full width = 20 km $\rm s^{-1}$) of the$\rm ^{12}$CO
$J$=2-1 line, further verify that there is an outflow motion. The
outflow emission is elongated along the north-SW  direction. This
region also is associated with an HII region and an IRAS source, but
we cannot identify which object is the driving source of the outflow
owing to our lower spatial resolution. An H$_{2}$O maser is also
associated with this region. Comparing with the components of the
outflow, we find that the H$_{2}$O maser is not associated with the
outflow, but may be associated with the core of  IRAS 20188+3928.
Using the LVG model, the possible kinetic temperature of the
blueshifted lobe and redshifted lobe are 26.9 K and 52.9 K,
respectively. The total gas mass, average dynamical timescale, and
mass entrainment rate of the outflow are 3.5 $\rm M_{\odot}$,  6.2
$\times$ $10^{4}$ yr and 6.2 $\times$ $10^{-5}$ $\rm M_{\odot}\
yr^{-1}$, respectively. The outflow has the higher integrated
intensity ratio ($I_{\rm CO}$$_ {J=3-2}$/$I_{\rm CO}$$_ {J=2-1}$),
which may be caused by the outflow interacting with the surrounding
ISM. The maximum ratio value of the redshifted lobe is greater than
that of the blueshifted lobe, we suggest that the redshifted lobe
has the higher temperature.

\begin{acknowledgements}
We thank anonymous referee for his/her constructive suggestions.
Jin-Long Xu's research is in part supported by{ the} 2011 Ministry
of Education doctoral academic prize{ and a}lso supported by the
young researcher grant of {N}ational {A}stronomical {O}bservatories,
Chinese {A}cademy of {S}ciences. Jin-Long Xu thanks Dr. Martin,
Miller and Ms. Ni-Mei Chen for their helps during the observations.
\end{acknowledgements}

\label{lastpage}

\end{document}